\documentclass[aps,prb,twocolumn,groupedaddress,amsmath,amssymb,showpacs]{revtex4}

\usepackage{color}

\usepackage{graphicx}
\usepackage{hyperref}
\usepackage{pifont}

\begin{document}
\title{Origin of the $n$-type and $p$-type conductivity of MoS$_2$ monolayers on a SiO$_2$ substrate}
\author{Kapildeb Dolui, Ivan Rungger and Stefano Sanvito}
\affiliation{School of Physics and CRANN, Trinity College, Dublin 2, Ireland}
\date{\today}
\begin{abstract}
\textit{Ab-initio} density functional theory calculations are performed to study the electronic properties of a MoS$_2$ 
monolayer deposited over a SiO$_2$ substrate in the presence of interface impurities and defects. When MoS$_2$ 
is placed on a defect-free substrate the oxide plays an insignificant role, since the conduction band top and the valence 
band minimum of MoS$_2$ are located approximately in the middle of the SiO$_2$ band-gap. However, if Na impurities 
and O dangling bonds are introduced at the SiO$_2$ surface, these lead to localized states, which modulate the conductivity 
of the MoS$_2$ monolayer from $n$- to $p$-type. Our results show that the conductive properties of MoS$_2$ deposited 
on SiO$_2$ are mainly determined by the detailed structure of the MoS$_2$/SiO$_2$ interface, and suggest that doping the
substrate can represent a viable strategy for engineering MoS$_2$-based devices.
\end{abstract}
\maketitle
\section{Introduction}

Recently MoS$_2$-based layered transition metal chalcogenides (LTMDs) have attracted considerable attention due to 
their potential for constructing low dimensional nano-structures for a variety of applications~\cite{Nature_2007_2_53, AcsNano_2012_6_74,JNC2012}. The electronic properties of MoS$_2$ show a strong dependence on 
thickness~\cite{NL_2010_10_1271}, i.e. on the number of atomic layers forming a given sample. In particular, 
MoS$_2$ monolayers, which display a substantial direct band-gap, represent a semiconducting alternative to 
graphene, which is a metal in its pristine form. Although the band-gap of graphene can be opened by fabricating 
nanoribbons~\cite{PRL_2008_100_206803} or by depositing it on a suitable substrate~\cite {Nature_2007_6_770}, 
this comes to the prize of deteriorating, in a somehow uncontrollable way, the carrier mobility due to edges and impurity 
scattering~\cite{PNAS_2007_104_18392}. In contrast, the low dimensionality, the small amount of dangling 
bonds, and their typical high crystalline form, make the performances of LTMD-based transistors comparable 
to those of existing Si-based ones~\cite{JAP_2007_101, IEEE_2011_58, NL_2012_12_1538}. In particular transistors 
made from MoS$_2$ monolayers have been recently fabricated, showing a mobility of at least 200~cm$^{2}$/V$\cdot$s 
at room temperature, an on/off current ratio of 10$^8$ and low standby power dissipation~\cite{Nature_2011_6}. 

Interestingly, both $n$-type~\cite{Nature_2011_6, PRL_2010_105, Adv_2012_24_2320, Small_2012_8_682} and 
$p$-type~\cite{ACIEd_2011_50_11093, Small_2012_8_966} conductivities have been reported in ultra-thin MoS$_2$
layers deposited on SiO$_2$. The conducting behavior of MoS$_2$ therefore seems to depend on the experimental 
details and an explanation for the specific current polarity ($n$- or $p$-type) remains far from being clear. Note that 
no intentional doping was introduced in the experiments mentioned above, so that the source of the different carrier 
types should be intrinsic to the MoS$_2$ layer, to the substrate and to the interaction between the two.

The possible creation of Mo and/or S vacancies during the growth cannot be the cause of the various conductive properties, 
since vacancies create deep levels at mid-gap in the bandstructure of MoS$_2$ monolayers \cite{JPChC_2011_115_13303}. 
Notably, disorder at the semiconductor/substrate interface in general plays a crucial role in determining the conductive properties 
of ultra-thin devices. For example, for GaAs nanowires it has been demonstrated that upon decreasing the nanowire diameter 
the interface-mediated conductivity gradually becomes dominant over the bulk one~\cite{AcsNano_2012_6_4428}. 
Such surface sensitivity can also be used to one's advantage. For instance an ambipolar transistor has been realized in 
MoS$_2$ thin flakes by contact with a ionic liquid environment~\cite{NL_2012_12_1136}, which as well affects the interface 
properties. Since MoS$_2$ monolayers are placed on insulators in practically any device architecture, it is important to identify 
the possible effects that the substrate has on the conductivity.

The defects responsible for the conductive properties of low dimensional devices are expected to be extrinsic in nature, 
such as charged impurities at the interface between the conductive channel and the substrate. These lead to an inhomogeneous 
Coulomb potential for both conduction and valence band electrons. Such charge traps have been identified to be in the form of 
adsorbates or defects at the surface of the underlying substrate in the case of 
graphene~\cite{JPCM_2010_22_334214, APL_2008_93_202110}. 
Likewise, temperature dependent transport measurements on thin MoS$_2$ layers, down to the monolayer limit, suggest that 
trapped charges at the SiO$_2$ surface could be responsible for the observed $n$-type behavior, when MoS$_2$ is deposited 
on SiO$_2$~\cite{AcsNano_2011_5_7707}. 

In general when charged traps are located at an interface, they influence the depletion/accumulation of electrons in the conducting 
channel up to a certain thickness, which is proportional to the channel screening length. This distance depends on different physical 
features, such as the nature and the density of the traps, and the electronic properties of the channel. For conventional semiconductors 
it typically reaches up to a few nanometers. For instance it has been recently demonstrated that charged trap states at the 
substrate/channel interface significantly affect the conductivity of GaAs nanowires up to diameters of about 
40-70~nm \cite{AcsNano_2012_6_4428}. More dramatic effects are expected for layered compounds down to the single layer limit,
in which essentially all the atoms are at the interface with the substrate and the channel vertical dimension is certainly shorter than the
screening length. Recently a reduction in conductivity with increasing the MoS$_2$ film thickness has been observed in MoS$_2$-based 
transistors, where SiO$_2$ was used as back gate \cite{IEEE_2012_33_1273}. This suggests that for the MoS$_2$/SiO$_2$ system the 
transport is interface-mediated, as intrinsic defects, homogeneously distributed in MoS$_2$, would not lead to any dependence of the 
conductivity on thickness. 

In order to shed some light on the effects that trap states at the SiO$_2$ surface have on the conductive properties of 
MoS$_2$/SiO$_2$ hybrid systems, we have performed state of the art first principle electronic structure calculations. In 
particular we have considered the case when the traps are due to impurities such as immobile Na and H atoms, and 
O-dangling bonds. 
The paper is organized as follows. In the next section we briefly describe our computational techniques and we provide details 
of the simulations performed. Then we proceed with presenting the results of this work in the context of recent experiments,
and finally we conclude.

\section{Methodology}
In order to investigate the influence of a SiO$_2$ substrate on the electronic properties of a MoS$_2$ monolayer, \textit{ab-initio} 
calculations are performed by using density functional theory (DFT)~\cite{PRB_1964_136, PRA_1965_140} within the generalized 
gradient approximation (GGA) of the exchange and correlation (XC) functional as introduced by Perdew, Burke and 
Ernzerhof (PBE) ~\cite{PRL_1996_77_3865} and numerically implemented in the SIESTA code \cite{IOP_2002_14}. In our 
calculations, a double-$\zeta$ polarized~\cite{PRB_2001_64} numerical atomic orbital basis set is used for all the atoms and 
the Troullier-Martins scheme is employed for constructing norm-conserving pseudopotentials~\cite{PRB_1991_43}. An equivalent 
plane wave cutoff of 350~Ry is chosen in all the simulations and the Brillouin zone is sampled by using an equivalent $k$-grid cutoff 
of 17~\AA. Relaxed geometries are obtained with the conjugate gradient method, where all the atoms in the supercell are allowed 
to relax until the force on each atom is less than 0.02~eV/\AA.

A trap state is usually formed when an energy level associated either to a defect or an impurity appears within the energy gap of the 
host material. Such trap states influence the charge transport properties mainly in two ways. Firstly, if the traps are charged, they will 
capture a hole or an electron from the environment. This produces a modification of the electrostatic potential, which in turns shifts the 
level alignments in the system, and thus affects the conductivity. Secondly, they can also increase the carrier concentration and provide 
pathways for electrons or holes to hop. The efficiency of this process depends on the amount of localization of the states associated 
to the defect site. If the energy of the localized gap state is close to either the valence band maximum (VBM) or the conduction band 
minimum (CBM), then at a given temperature some of these charges will be transferred either to the conduction or to the valence band, 
where they may contribute to increase the system conductivity. 

Whether or not one can describe with {\it ab initio} calculations such mechanisms depends crucially on the ability of computing 
accurately the energy levels of the system. The use of the GGA (or of the local density approximation - LDA) for electronic structure 
calculations of defect levels is, in general, problematic. One reason is the typical underestimation of the energy gap and the related 
incorrect alignment of the energy levels of hybrid systems. For instance an artificially reduced band-gap may erroneously bring deep 
traps in resonance with either the conduction or the valence band~\cite{DasComment,ZungerPhysics}. A second source of error is 
the incorrect description of the charge localization at the defect site, a feature that usually leads to predict defects to be too 
shallow~\cite{Andrea}. Atomic self-interaction correction (ASIC)~\cite{PRB_2007_75_045101,VPSIC} has been proved to overcome 
these deficiencies~\cite{Andrea2,DasZnO}. Therefore we also perform additional LDA+ASIC calculations to verify the robustness of the 
LDA/GGA results. In particular we set the ASIC scaling parameter to $\alpha=0.5$, a value which is generally appropriate to mid-gap
insulators~\cite{PRB_2007_75_045101}.

\section{Results and discussion}

\subsection{Defect-free SiO$_2$ interface}
Substantial experimental efforts have been devoted to deposit ultra-thin MoS$_2$ layers onto SiO$_2$ in order to demonstrate 
transistor operation, down to the single layer limit~\cite{Nature_2011_6, PRL_2010_105, Small_2012_8_966}. Usually amorphous 
oxides are used as substrates. However, in order to avoid the computational complexity of a highly disordered structure, a crystalline 
SiO$_2$ substrate is simulated here instead. This also allows us to systematically determine the effects of individual defects and 
impurities on the electronic structure of a MoS$_2$ layer. Our unit cell is constructed as a slab containing at least 6 Si atomic layers 
of $\alpha$-quartz and an adsorbed MoS$_2$ monolayer. At least 15~\AA~of vacuum are included at the slab boundaries to avoid 
the spurious interaction between the slab periodic images. We consider the modified oxygen-terminated (0001) SiO$_2$ surface in 
order to simulate the most experimentally relevant conditions. 

\begin{figure}
\center
\includegraphics[width=6.0cm, clip=true]{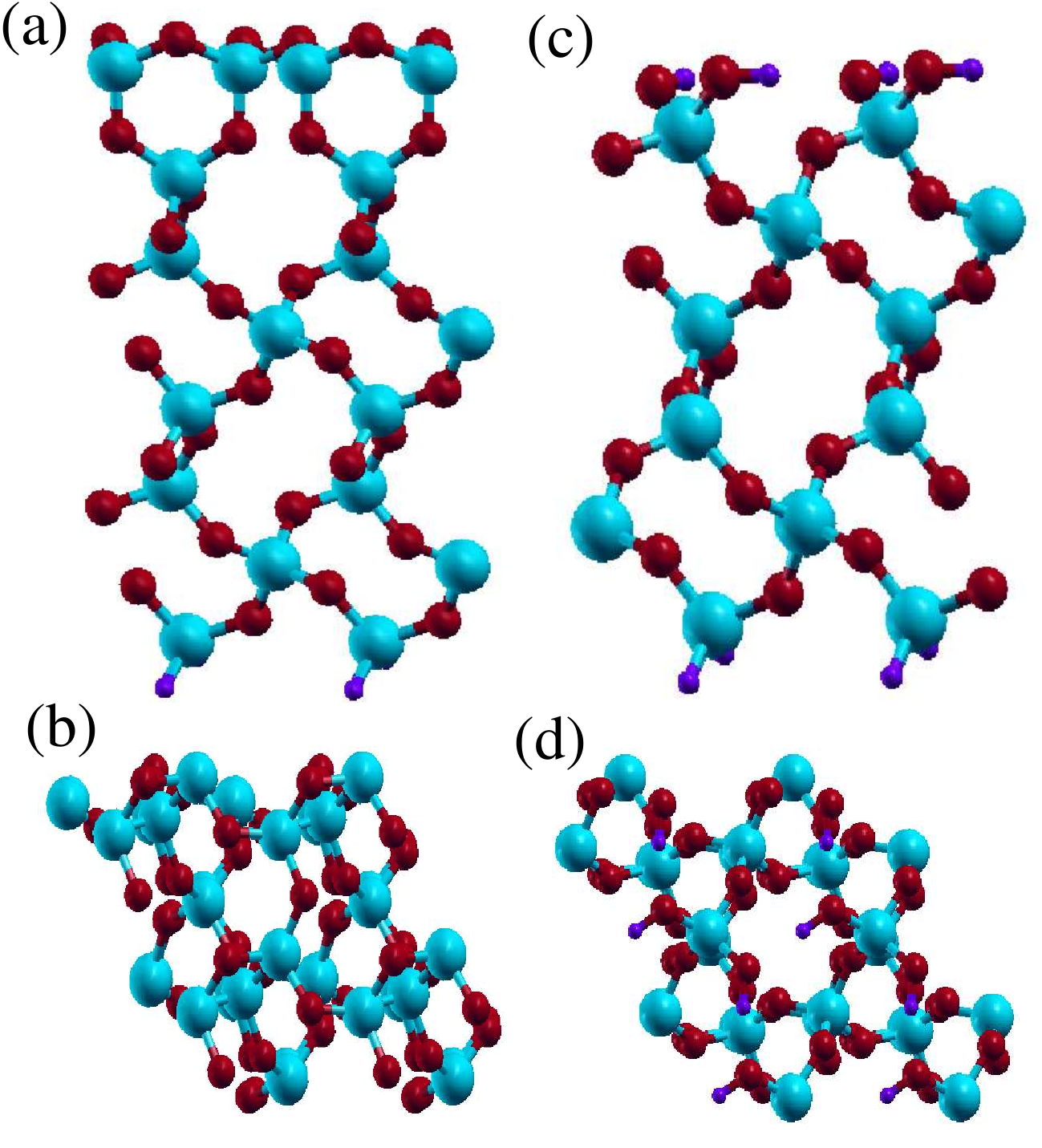}
\caption{(Color online) Side and top views of reconstructed structures for the O-terminated SiO$_2$ (0001) surface (a and b), and the 
fully hydroxylated SiO$_2$ (0001) one (c and d) (color code: cyan $\rightarrow$ Si, red $\rightarrow$ O, violet $\rightarrow$ H).}
\label{fig:SiO2}
\end{figure}

Two primary structures for the oxygen-terminated SiO$_2$ (0001) surface are possible, depending on whether the termination is
with the siloxane group (Si-O-Si) or with the silanol one (Si-OH). Both surfaces can form depending on the surface 
treatment~\cite{JAP_2011_110_024513}. The siloxane reconstruction at room temperature forms an O-terminated surface with 
an outermost six-membered ring structure, as shown in Fig.~\ref{fig:SiO2}(a, b). Under annealing in ambient conditions it becomes
hydroxylated (Si-OH) and the reconstruction transforms into the silanol one, which presents on the surface a zigzag H-bonded network 
[see Fig.~\ref{fig:SiO2}(c, d)]. In both cases in our simulations the dangling bonds on the Si-terminated bottom surface are saturated 
by hydrogen. 

\begin{figure}
\center
\includegraphics[width=8.0cm,clip=true]{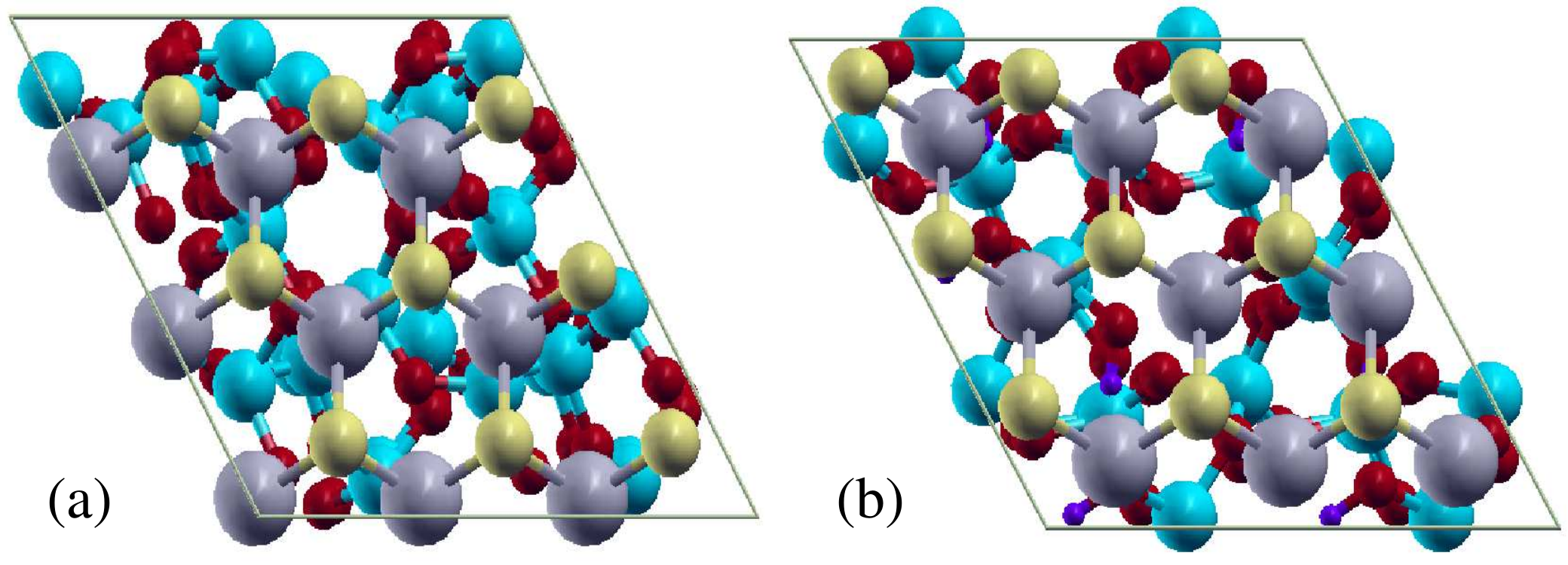}
\caption{(Color online) Top view of the optimized structure of MoS$_2$ placed on a defect-free (a) siloxane and (b) silanol surface 
(color code: light grey $\rightarrow$ Mo, yellow $\rightarrow$ S, cyan $\rightarrow$ Si, red $\rightarrow$ O, violet $\rightarrow$ H).}
\label{fig:SiO2/MoS2}
\end{figure}

The optimized lattice constants of the pristine SiO$_2$ and MoS$_2$ are 4.91~\AA~and 3.19~\AA, respectively. We therefore construct 
a hexagonal supercell in the plane, with a 9.69~\AA-long side, so that the lattice mismatch between SiO$_2$ and MoS$_2$ is minimized 
to $\sim$1.2~\%. The GGA calculated band-gap of SiO$_2$ and of a MoS$_2$ monolayer are 6.20~eV and 1.49~eV, respectively. The 
small strain applied to the MoS$_2$ monolayer changes only little the electronic structure. The band-gap remains direct at the K point 
and it is only reduced by 0.22~eV from the value of 1.71~eV obtained for the unstrained case. Similarly to the case of graphene 
\cite{PRL_2011_106_106801}, we expect that the electronic structure of a MoS$_2$ monolayer is only marginally affected by its 
local arrangement on the SiO$_2$ substrate. Therefore, as a representative configuration we use the arrangement of 
Fig.~\ref{fig:SiO2/MoS2}, where an oxygen atom is situated at the hollow site of the Mo surface triangles.

We start our discussion by presenting the properties of the defect-free hybrid MoS$_2$/SiO$_2$ system. The equilibrium distances, 
$d_0$, between the SiO$_2$ and the MoS$_2$ surfaces are 3.01~\AA~and 2.98~\AA~for siloxane and silanol,~respectively.
Here $d_0$ is defined as the vertical separation between the top-most O layer in the SiO$_2$ surface and its nearest S layer in MoS$_2$.
These values are similar to the distance between two MoS$_2$ monolayers that we calculate to be 3.17~\AA. The binding energy of the 
MoS$_2$/SiO$_2$ system is given by $E_\mathrm{b}$ = $E_\mathrm{MoS_2}$ + $E_\mathrm{SiO_2}$ - $E_\mathrm{MoS_2/SiO_2}$, 
where $E_\mathrm{MoS_2}$, $E_\mathrm{SiO_2}$, and $E_\mathrm{MoS_2/SiO_2}$ are total energies of the isolated MoS$_2$, 
the isolated SiO$_2$ slab, and of the MoS$_2$/SiO$_2$ hybrid system, respectively. We find $E_\mathrm{b}$ for siloxane and silanol 
to be respectively 0.14~eV and 0.16~eV per primitive MoS$_2$ unit cell. These binding energies are close to that between two 
MoS$_2$ layers (0.20 eV/unit cell), which are bound together by the rather weak van der Waals forces. As such, our results show that 
MoS$_2$ is weakly bound also to the SiO$_2$ surface, in agreement with recent experimental results that have measured the 
interaction between MoS$_2$ and an underlying SiO$_2$ substrate to be negligible~\cite{PRB_2011_84_045409}. 

Note that in general GGA-type XC-functionals do not describe accurately van der Waals forces. However, it has been shown that the 
LDA/GGA is able to reproduce the interlayer spacing and the binding energy of layered chalcogenides~\cite{PRL_2012_108_235502}. 
We have then verified that our calculated $d_0$ for bulk MoS$_2$, $d_0=3.08$~\AA, is in good agreement with the experimental value 
of 2.96~\AA~\cite{PRB_1976_13_3843} and also with the previously calculated theoretical estimate of 
3.05~\AA~\cite{JPChC_2011_115_16354}. Moreover, in order to take into account possible small deviations of the relaxed distance 
from the experimental value due to the XC functional used, we have evaluated the electronic structure for $d_0$ within the range 
$d_0~\pm$0.5~\AA, and we have found that the results change little with varying the distance. 

\begin{figure}
\center
\includegraphics[width=8.0cm,clip=true]{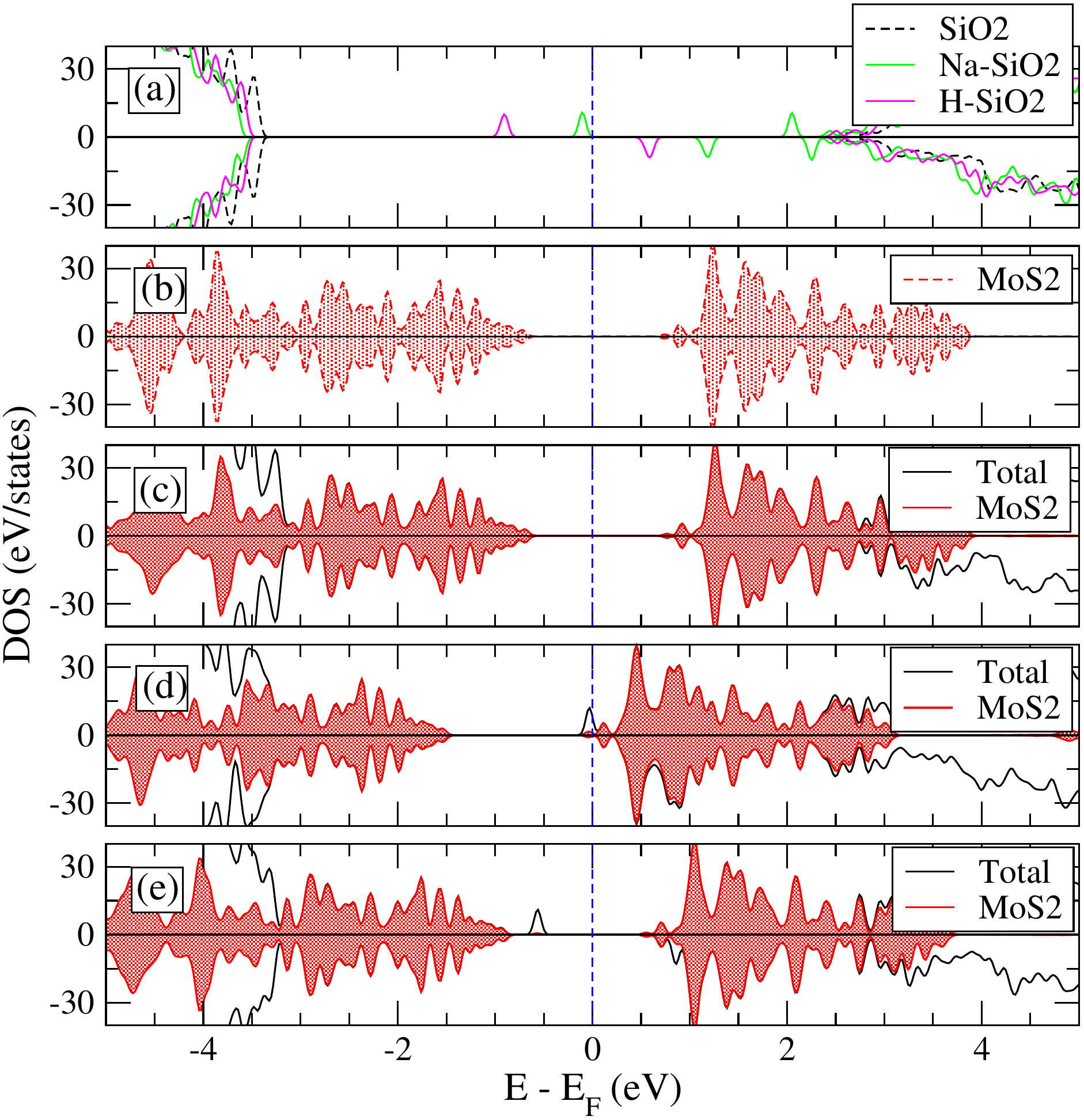}\caption{(Color online) Electronic structure of the SiO$_2$/MoS$_2$ hybrid system when 
various defects are present at the SiO$_2$ siloxane surface. (a) The total DOS for the defect-free surface (black, dashed curves),
and when either Na (green, solid curves), or H adsorbed (magenta, solid curves) are adsorbed. (b) The DOS of a pristine free-standing
MoS$_2$ monolayer. The total DOS and the PDOS for MoS$_2$, when the MoS$_2$ monolayer is placed on the defect-free siloxane 
surface (c), on a siloxane surface with one adsorbed Na, or (d) with one adsorbed H. The blue dashed vertical line indicates the Fermi 
level, which has been set to zero in all the panels. The red shaded areas indicate the MoS$_2$ PDOS. Positive and negative DOS 
are respectively for spin up (majority spins) and spin down (minority spins) electrons.}
\label{fig:DOS1}
\end{figure}
\begin{figure}
\center
\includegraphics[width=8.0cm,clip=true]{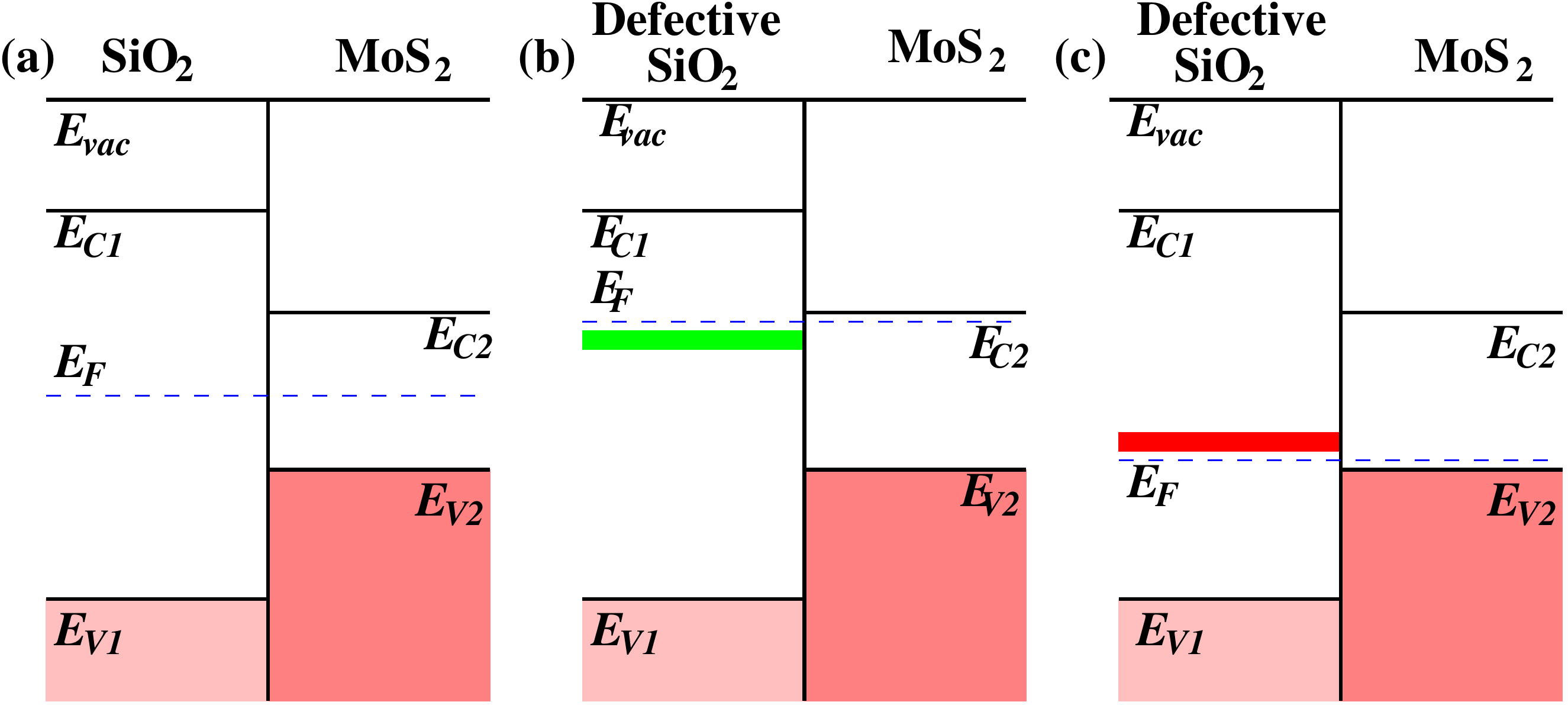}
\caption{(Color online) Schematic band diagram for MoS$_2$ placed on the defect SiO$_2$ substrate (a), and on a substrate including 
a defect-induced donor (b) or acceptor (c) level. This demonstrates the modulation of the conductivity from $n$-type to $p$-type as 
the impurity state redefines the Fermi energy in the oxide. The energy levels $E_{V}$, $E{_C}$, $E_F$ and $E_\mathrm{vac}$ define 
the valence band maximum (VBM), the conduction band minimum (CBM), the Fermi energy, and the vacuum level, respectively. 
The subscripts 1 and 2 refer to SiO$_2$ and MoS$_2$, respectively. The blue dashed-lines indicate the Fermi energy of the hybrid 
system (common to SiO$_2$ and MoS$_2$). The thick green line in (b) indicates the donor level and the thick red line in (c) represents 
the acceptor state in the oxide. Note that in general due to charge transfer from MoS$_2$ to the gap states, and the related dipole 
formation, the level alignment between $E_{V1}$ and $E_{V2}$ will also change in the defective systems.}
\label{fig:schematic}
\end{figure}

\subsection{SiO$_2$/MoS$_2$ composite with siloxane reconstruction}
We now move to study the electronic structure of a MoS$_2$ monolayer deposited on SiO$_2$ by starting with the siloxane surface. 
In particular we consider first the situation where SiO$_2$ is defect-free. Fig.~\ref{fig:DOS1}(c) shows the density of states (DOS) of 
the hybrid SiO$_2$/MoS$_2$ system, which remains semiconducting with a band-gap of 1.48~eV, i.e. with the same band-gap of a 
free-standing MoS$_2$ monolayer having the same lattice parameters. Both the valence and the conduction bands of the hybrid 
compound are associated to MoS$_2$. We note that the projected DOS (PDOS) over MoS$_2$ extends into the SiO$_2$ band-gap, 
and the total DOS of the combined material is essentially given by the superposition of the DOSs of the pristine slab of SiO$_2$ 
[Fig. \ref{fig:DOS1}(a)] and of the MoS$_2$ monolayer [Fig.~\ref{fig:DOS1}(b)]. Both the conduction and the valence bands of SiO$_2$ 
are located at least 1.5~eV away from those of MoS$_2$. As a consequence, no charge transfer between the substrate and 
MoS$_2$ occurs. Importantly, one of the basic criteria for the selection of the gate oxide is fulfilled here, namely that the oxide should 
have a bands offset of over 1~eV for both the conduction and valence band in order to create a large barrier for both electrons and 
holes~\cite{JAP_2006_100_014111}. Our results show that the conductivity of MoS$_2$ is not influenced by the underlying defect-free 
SiO$_2$ substrate. Therefore the measured $n$-type or $p$-type conducting properties of MoS$_2$ on SiO$_2$ must be due to defects 
and impurities.

Localized states, arising from impurities or defects within the oxide substrate or at the interface with the conducting channel, can 
redefine the effective Fermi level of the hybrid system, as illustrated schematically in Fig.~\ref{fig:schematic}. Depending on the 
alignment of the gap states with respect 
to the MoS$_2$ valence and conduction bands, the system can switch from $n$-type [see Fig.~\ref{fig:schematic}(b)] to $p$-type [see 
Fig.~\ref{fig:schematic}(c)]. Therefore, such trap states are expected to give significant contributions to the conductivity of these low 
dimensional systems. In the layered structure considered in this work, the trap states are expected to be located at the interface between 
the LTMDs and the substrate, not in the LTMDs themselves, which usually are highly defect-free. Trap states at the SiO$_2$ surface 
can have a wide range of origins, such as immobile ionic charges, SiO$_2$ surface dangling bonds, and foreign impurities adsorbed 
on the surface~\cite{book_1}. In literature densities of trap states on SiO$_2$ are reported in the range~\cite{TrapDensity}~
10$^{10}$-10$^{14}$ cm$^{-2}$. As representative dopants, here we consider two possible candidates: Na atoms and SiO$_2$ 
surface oxygen dangling bonds.
\begin{figure}
\center
\includegraphics[width=8.0cm,clip=true]{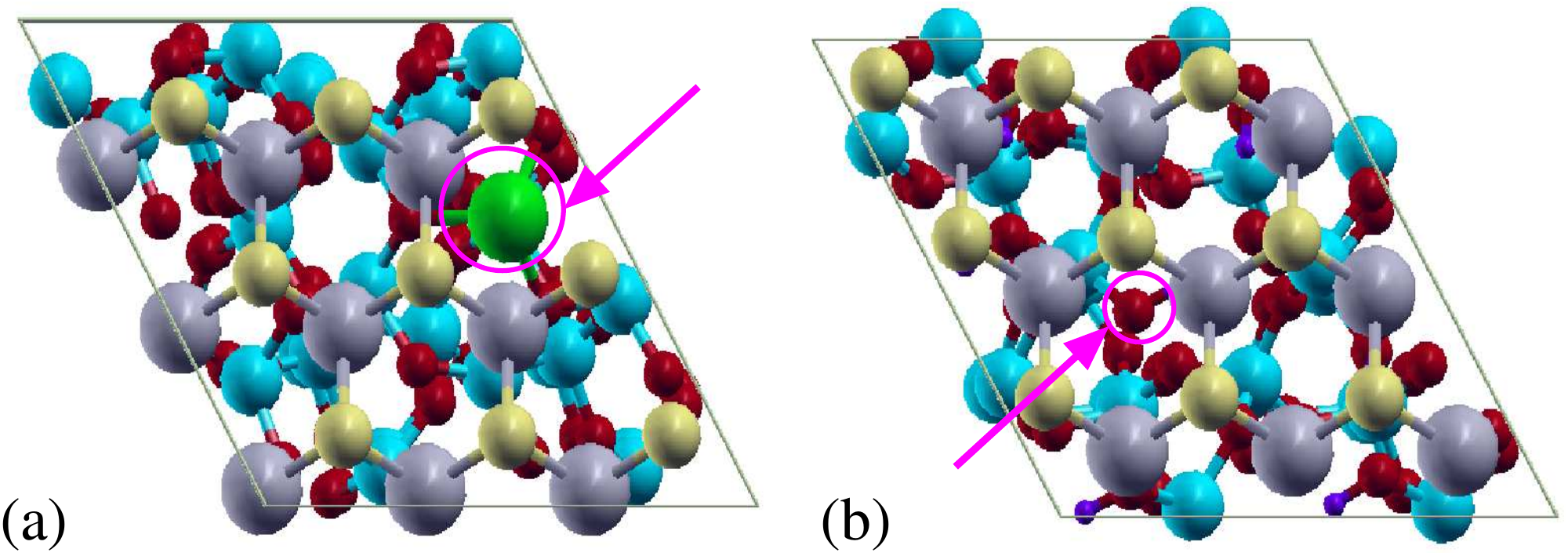}
\caption{(Color online) The optimized geometry for MoS$_2$ placed on (a) the siloxane surface incorporating a Na impurity and (b) 
the dangling oxygen bond on the silanol surface, obtained by removing a H atom. The arrows indicate the positions of the defects on 
the surface (Color code: green $\rightarrow$ Na, while the other colors are the same atoms as in Fig.~\ref{fig:SiO2/MoS2}). 
The arrows indicate the location of the impurities/defects.} 
\label{fig:Trap_SiO2/MoS2}
\end{figure}

During the synthesis and the sample preparation, SiO$_2$ can adsorb relatively light impurities such as Na and K~\cite{book_2} at
its surface. In order to simulate the effects of such impurities on the electronic structure of the MoS$_2$ channel a single Na atom is 
placed on top of the siloxane SiO$_2$ surface. Given the lateral dimension of our supercell, this corresponds to an impurity density 
of $\sim$10$^{14}$ cm$^{-2}$, which is close enough to the recently reported values of trap states densities, reaching up to 
$\sim$10$^{13}$ cm$^{-2}$ for thin MoS$_2$ layers deposited on SiO$_2$~\cite{JAP_2007_101, IEEE_2012_33_1273}. 
The most energetically favorable binding position for Na is found to be at the center of the surface oxygen triangle [see panel (a) 
of Fig.~\ref{fig:Trap_SiO2/MoS2}]. A Na adatom adsorbed on a pristine SiO$_2$ surface creates a deep donor state in the DOS, 
with a single particle level at about 2~eV below the SiO$_2$ CBM [see Fig.~\ref{fig:DOS1}(a)]. Note that such state is singly occupied
and therefore spin-splits in our spin-polarized calculations, with the empty minority spin state (spin down) laying approximately 1~eV 
below the CBM and 1~eV below the Fermi level. 

When a MoS$_2$ monolayer is deposited over the the Na-doped SiO$_2$ surface, $d_0$ increases to 3.24~\AA\ at the edges of 
our unit cell, whereas at the Na site the O-S distance becomes 3.45~\AA. The enlargement of the binding distance compared to that
of the pristine SiO$_2$/MoS$_2$ system is a direct consequence of the Na intercalation at the interface. The electronic structure of 
the composite is strongly affected by the presence of the Na ion, as shown in Fig.~\ref{fig:DOS1}(c). Also in this case the total DOS
appears as a direct superposition of those of SiO$_2$ and MoS$_2$. However the presence of the Na filled state shifts the Fermi level, 
which now gets pinned just below the MoS$_2$ CBM. The resulting DOS around $E_\mathrm{F}$ is thus that of the defect-free 
MoS$_2$ conduction band with the addition of a Na-derived impurity level positioned below it. Hence, the gap state is moved below the 
Fermi energy, resulting in a very small activation energy for the transfer of electrons from Na to the MoS$_2$ conduction band. This is the 
situation schematically presented in figure~\ref{fig:schematic}(b), which leads to $n$-doping. 

If we now replace Na with H on the SiO$_2$ siloxane surface, the associated filled gap state lies deep in the SiO$_2$ band-gap 
[see Fig.~\ref{fig:DOS1}(a)], despite the fact that H and Na share the same $s$-like valence. The same situation persists in the 
composite [see Fig.~\ref{fig:DOS1}(e)], where the H-derived filled spin-up level remains at mid-gap, approximately 0.5~eV above 
the VBM, while the empty spin-down one is nearly resonant within the conduction band. This situation however does now lead to 
doping so that H can not influence the conductivity of the MoS$_2$/SiO$_2$ structure. The quantitative difference found between the 
results for the Na and the H case show that, in order to obtain $n$-type character, only impurities with rather small ionization potential 
are relevant. These can transfer one electron to the MoS$_2$ conduction band with small activation energy. Such activation energy is 
a key factor in the determination of the threshold voltage, $V_\mathrm{th}$, required to operate a transistor in the on-state. As a
consequence, the experimentally measured values for $V_\mathrm{th}$, which show a large variation for different samples 
\cite{AcsNano_2011_5_7707}, are then attributable to varying concentrations and properties of the trap states from sample to sample.
\begin{figure}
\center
\includegraphics[width=8.0cm,clip=true]{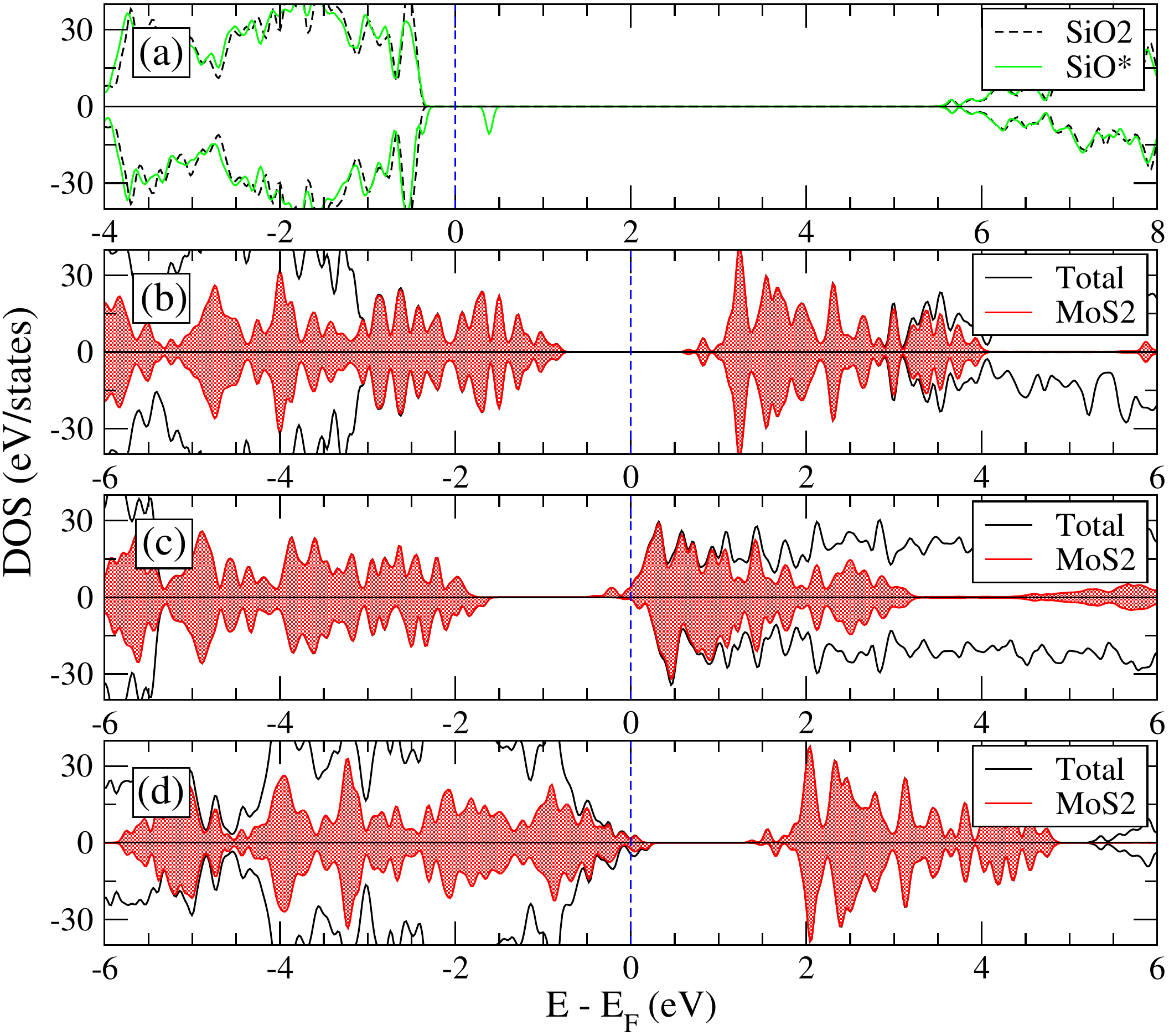}
\caption{(Color online) Electronic structure of the SiO$_2$/MoS$_2$ hybrid system when various defects are present at the SiO$_2$ 
silanol-reconstructed  surface. (a) The DOS for the defect-free surface (black, dashed curve) and for the one where one O dangling 
bond is induced by a single H removal (green, solid curve labelled as SiO$^*$). The total DOS and the MoS$_2$ PDOS for the 
SiO$_2$/MoS$_2$ composite when the MoS$_2$ monolayer is placed on (b) the defect-free surface, (c) on the surface with a 
single adsorbed Na atom, and (d) on the surface with a O dangling bond created by removing a single H atom. The blue dashed line 
indicates the Fermi energy, which is set to zero in all panels. The red shaded areas indicate the MoS$_2$ PDOS. Positive and negative 
DOS are respectively for spin up (majority spins) and spin down (minority spins) electrons.}
\label{fig:DOS2}
\end{figure}

\subsection{SiO$_2$/MoS$_2$ composite with silanol reconstruction}
Next we move to examine the case of the SiO$_2$ surface with silanol reconstruction, whose DOS is presented in Fig.~\ref{fig:DOS2}(a).
Similarly to the siloxane case, the PDOS for the defect-free MoS$_2$/SiO$_2$ composite [see Fig. \ref{fig:SiO2}(b)] corresponds to a
superposition of the DOSs of the isolated MoS$_2$ [Fig. \ref{fig:DOS1}(b)] and SiO$_2$ [Fig.~\ref{fig:DOS2}(a)] components, 
indicating weak interaction between the two materials. When a Na atom is intercalated between the silanol surface and the MoS$_2$
layer, we find that the system becomes $n$-type [Fig.~\ref{fig:DOS2}(c)], in the same way as for the siloxane surface. This indicates
that Na is an efficient $n$-dopant for MoS$_2$ on SiO$_2$ regardless of the surface reconstruction.

In general thermal annealing of the silanol surface creates under-coordinated oxygen atoms (Si-O*). These appear as stable surface
defect centers and act as typical charge traps in oxygen rich SiO$_2$ \cite{JNCrysS_1979_32}, since they are able to capture an extra 
electron in their dangling bond. In our calculations, such defects are created on the Si-OH surface by removing a H atom from the top 
surface [see Fig.~\ref{fig:Trap_SiO2/MoS2}(b)]. For such a defect we find that the empty acceptor state is created $\sim$0.9~eV above 
the SiO$_2$ VBM [see Fig.~\ref{fig:DOS2}(a)]. Once MoS$_2$ is layered onto the surface, the value of $d_0$ at the boundary of our 
H-deficient unit cell is $d_0=2.98$~\AA, which is approximately equal to that for the pristine surface, whereas at the dangling bond site 
the O-S distance is significantly reduced to 2.68 \AA. When placing the MoS$_2$ monolayer on this defective surface, the dangling
bond state gets filled by capturing an electron from the MoS$_2$ valence band, so that the Fermi energy now lies just below the MoS$_2$ 
VBM [see Fig.~\ref{fig:DOS2}(d)]. This is the level alignment presented in Fig.~\ref{fig:schematic}(c), which makes the composite 
$p$-type. Note that the rather high density of oxygen dangling bonds in our system causes a large surface charge density dipole, 
which shifts the MoS$_2$ DOS downwards in energy by more than 1~eV with respect to the SiO$_2$ substrate. By modulating the 
density of such defect types one may be able to change such a shift.

\subsection{Robustness of the results against the choice of XC functional: ASIC}
\begin{figure}
\center
\includegraphics[width=8.0cm,clip=true]{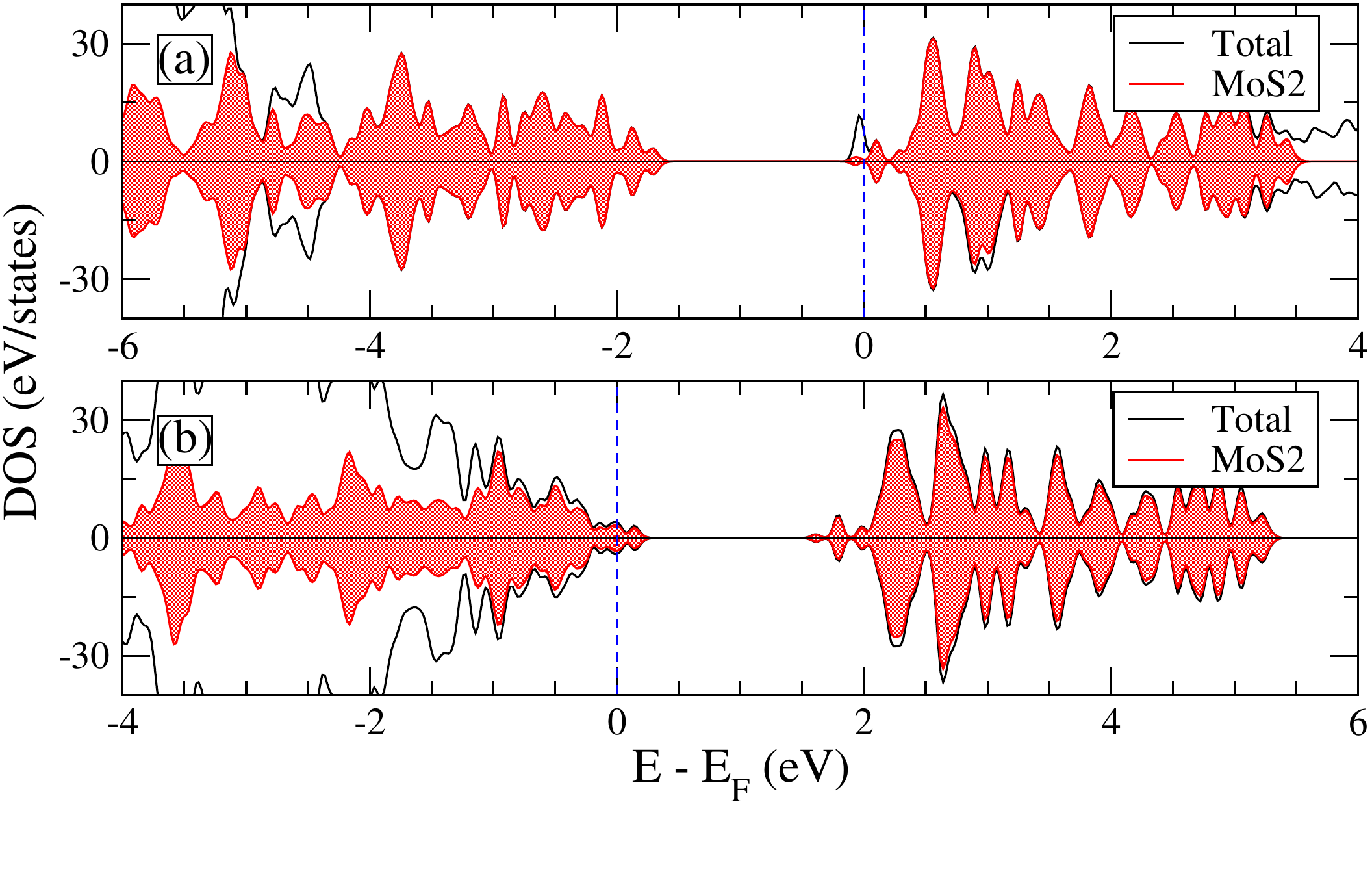}
\caption{(Color online) Density of states for the defective SiO$_2$/MoS$_2$ composite calculated with the ASIC XC functional.
In panel (a) we report the DOS for the siloxane reconstruction with an intercalated Na atom [corresponding to Fig~\ref{fig:DOS1}(d)],
while in (b) that for the silanol reconstruction and an O dangling bond obtained by removing a surface H atom 
[corresponding to \ref{fig:DOS2}(d)].}
\label{fig:SIC}
\end{figure}
Finally, in order to verify that the calculated level alignment is robust against the choice of exchange and correlation functional, we 
have repeated our calculations by using the ASIC scheme. As expected the ASIC functional increases the band-gap of 
MoS$_2$ and SiO$_2$ respectively to 1.73~eV and 8.02~eV (for the same strained hybrid structure used in the previous sections). 
In the case of SiO$_2$ this brings the calculated value sensibly closer to the experimental one of 8.9~eV \cite{SiO2BG}, as expected
from the ASIC when dealing with an insulator whose valence and conduction bands have different orbital 
content~\cite{PRB_2007_75_045101,VPSIC}.

The situation for MoS$_2$ is more complicated and deserves a detailed discussion. In this case the band-gap is defined by bands 
dominated mainly by Mo-$d$ orbitals and the ASIC opens it only marginally. For a free-standing MoS$_2$ monolayer the 
ASIC ($\alpha=0.5$) returns a direct band-gap of 2.03~eV (compared to a GGA gap of 1.71~eV). Note that the LDA value is 
1.87~eV so that the LDA already partially opens the gap with respect to the GGA. It is also notable that an enhancement of the screening 
parameter $\alpha$ to $\alpha=1$ (full atomic correction) produces a marginal further increase of the gap to 2.10~eV. Importantly the ASIC 
result is rather close to that calculated~\cite{Ramasubramaniam} with the hybrid Heyd-Scuseria-Ernzerhof (HSE) exchange-correlation 
functional~\cite{HSE}. 
This is however larger than the optical band-gap of 1.90~eV measured experimentally for MoS$_2$ monolayers~\cite{PRL_2010_105}. 
The apparent contradiction can be solved by noting that the optical excitations involve excitons with a large binding energy of the order
of 1~eV, as confirmed by many-body calculations~\cite{Ramasubramaniam}. Thus, one expects that the true quasi-particle 
spectrum has a band-gap of approximately 1.9+1=2.9 eV, in good agreement with that computed with the $GW$ scheme, either at the 
first order level~\cite{Ramasubramaniam} (2.82~eV) or self-consistently~\cite{Lamb} (2.76~eV). As such, the ASIC describes MoS$_2$ 
with a band-gap larger than the measured optical one and it provides an improved description over that of the GGA. 

We now go back to the SiO$_2$/MoS$_2$ composite and in Fig.~\ref{fig:SIC} we report two representative results for the case 
of Na adsorbed on the siloxane surface and for that of the O dangling bond on the silanol one. We find that for the first case, although 
the band-gaps of the two parental materials are both increased, the Fermi energy is still pinned at the bottom of the MoS$_2$ 
conduction band [Fig.~\ref{fig:SIC}(a)]. As a consequence Na still leads to a $n$-type semiconducting character with small activation 
barrier. Similarly the O dangling bond on the silanol-terminated surface leads to a $p$-type semiconducting character [see 
Fig.~\ref{fig:SIC}(b)], with the Fermi energy positioned below the MoS$_2$ valence band. This indicates that our two main results 
remain unchanged whether calculated at the GGA or ASIC level, i.e. they are robust with respect to the choice of exchange-correlation functional. 

\section{Conclusion}

The effects of the SiO$_2$ substrate on the conductivity of a semiconducting MoS$_2$ monolayer are investigated with first 
principles density functional theory calculations. The defect-free SiO$_2$ surface does not affect significantly the electronic 
properties of MoS$_2$ due to their weak mutual interaction. As such the conductive properties of MoS$_2$ do not change and
SiO$_2$ appear as an ideal gate material. However, when Na atoms are placed at the SiO$_2$/MoS$_2$ interface, a shallow 
donor trap state is created just below the CBM of the hybrid SiO$_2$/MoS$_2$ composite. The small activation energy makes 
the hybrid MoS$_2$/Na-SiO$_2$ system a $n$-type semiconductor even for rather low temperatures. Interestingly, the behavior 
is different for H adsorption, where the impurity level is created $\sim$0.9 eV below the CBM, resulting in a stable localized charge 
that cannot be easily promoted to the CBM and does therefore not affect the conductivity. 

In contrast, in the case of oxygen dangling bonds on the silanol-terminated SiO$_2$ surface, the Fermi energy of the 
MoS$_2$/SiO$_2$ system is located just below the VBM, making the system a $p$-type semiconductor. These results shows that 
the conductivity of ultra-thin semiconducting LTMDs changes from $n$-type to $p$-type depending on the charge-polarity of the 
traps, as well the energy level alignment of the trap states within the LTMDs band gap. These kind of trap states at the SiO$_2$ 
surface are likely to be at the origin of the observed change in conductance in different experimentally realized MoS$_2$-based 
transistors. Intriguingly, our results suggest the possibility of intentionally doping MoS$_2$ by depositing different adsorbates
over the substrate SiO$_2$ surface. This can pave the way for a new strategy in the design of two-dimensional devices,
where the electronic properties of the channel are engineered by manipulating those of the substrate.

\section*{Acknowledgments} This work is supported by Science Foundation of Ireland (Grant No. 07/IN.1/I945) and by CRANN. 
IR acknowledges financial support from the King Abdullah University of Science and Technology ({\sc acrab} project). 
We thank Trinity Centre for High Performance Computing (TCHPC) 
for the computational resources provided.


\end{document}